\documentclass{pasj00}
\draft
\begin{document}
\SetRunningHead{N. Isobe et al.}
      {Discovery of Suzaku J1305-4931}
\Received{yyyy/mm/dd}
\Accepted{yyyy/mm/dd}

\title{Discovery of a bright transient ultraluminous X-ray source
Suzaku J1305-4931 in NGC 4945}

\author{%
Naoki       \textsc{Isobe},      \altaffilmark{1}
Aya         \textsc{Kubota},     \altaffilmark{1,2}
Kazuo       \textsc{Makishima},  \altaffilmark{1,3}
Poshak      \textsc{Gandhi},      \altaffilmark{1}\\ 
Richard E.  \textsc{Griffiths},  \altaffilmark{4}
Gulab   C.  \textsc{Dewangan},  \altaffilmark{4}
Takeshi     \textsc{Itoh},       \altaffilmark{3}
\and 
Tsunuefumi    \textsc{Mizuno}     \altaffilmark{5}
}
\altaffiltext{1}{Cosmic Radiation Laboratory,
   the Institute of Physical and Chemical Research, \\
   2-1 Hirosawa, Wako, Saitama, 351-0198, Japan}
\email{isobe@crab.riken.jp}

\altaffiltext{2}{Department of Electronic Information Systems, 
Shibaura Institute of Technology, \\
307 Fukasakum Minuma-ku, Saitama-shi, Saitama, 337-8570, Japan}

\altaffiltext{3}{Department of Physics, University of Tokyo,
   7-3-1 Hongo, Bunkyo-ku, Tokyo, Japan}

\altaffiltext{4}{Department of Physics,
   Carnegie Mellon University, \\ 
5000 Forbes Avenue, Pittsburgh, PA 15213, USA}

\altaffiltext{5}{Department of Physics, Hiroshima University, \\
   1-3-1 Kagamiyama, Higashi-Hiroshima, Hiroshima 739-8526, Japan}

\maketitle
\begin{abstract}
This paper reports the discovery of a bright X-ray transient source,
Suzaku J1305-4913, 
in the south-west arm of the nearby Seyfert II galaxy NGC 4945. 
It was detected at a 0.5 -- 10 keV flux of
$2.2 \times 10^{-12}$ erg cm$^{-2}$ s$^{-1}$ 
during the Suzaku observation conducted on 2006 January  15 -- 17, 
but was undetectable in a shorter observation on 2005 August 22 --23, 
with an upper limit of $1.7 \times 10^{-14}$ erg cm$^{-2}$ s$^{-1}$
(90\% confidence level).
At a distance of $3.7$ Mpc, 
the bolometric luminosity of the source becomes 
$L_{\rm bol} = 4.4 \times 10^{39} \alpha$ erg s$^{-1}$, 
where $\alpha = (\cos 60^\circ / \cos i)$ and $i$ 
is the disk inclination. 
Therefore, the source is classified into so-called 
ultraluminous X-ray sources (ULXs).  
The time-averaged X-ray spectrum of the source is described by 
a multi-color disk model, 
with the innermost accretion disk temperature 
of $T_{\rm in} = 1.69_{-0.05}^{+0.06}$ keV.
During the 2006 January observation, 
it varied by a factor of $2$ in intensity, 
following a clear correlation of  $L_{\rm bol} \propto T_{\rm in}^4$. 
It is inferred that the innermost disk radius $R_{\rm in}$ stayed constant 
at $R_{\rm in} = 79_{-3.9}^{+4.0} \alpha^{1/2}$ km,
suggesting the presence of a standard accretion disk. 
Relating $R_{\rm in}$ with the last stable orbit 
around a non-rotating black hole yields a rather low black hole mass,
$\sim 9 \alpha^{1/2}$ solar masses,
which would imply that the source is shining at a luminosity
of $\sim3 \alpha^{1/2} $ times the Eddington limit. 
These results can be better interpreted 
by invoking sub-Eddington emission 
from a rapidly spinning black hole 
with a mass of $20$ -- $130$ solar masses.  
\end{abstract}

\section{Introduction}
\label{sec:intro}
A number of nearby galaxies host
bright off-center point-like X-ray sources,
with luminosities of $10^{39}$ -- $10^{40}$ erg s$^{-1}$ 
(e.g., \cite{X-ray_from_galaxies}).
These X-ray sources are usually called ultraluminous X-ray sources 
(ULXs; \cite{ULX_asca1}).
Based on their high luminosities,
ULXs are regarded as candidates for intermediate-mass black holes 
with masses of $M = 20$ -- $100 M_{\odot}$.
X-ray spectra of ULXs are usually described 
by a multi-color disk (MCD; \cite{MCD1,MCD2}) model, 
or a power-law (PL) model.
Because the MCD model provides a good approximation 
for the X-ray emission
from a standard accretion disk \citep{standard_disk}
and successfully interprets 
the observed X-ray spectra of Galactic black hole candidates,
the black hole scenario for ULXs seems plausible. 
However, a simple application of the MCD model to ULXs 
gives an innermost disk temperature 
($T_{\rm in} = 1$ -- $2$ keV, \cite{ULX_asca1}) significantly higher 
than those of Galactic black holes ($T_{\rm in} = 0.5$ -- $1$ keV).
This is apparently inconsistent with the argument 
for a intermediate-mass black hole,
since a more massive object would have a lower disk temperature
as $T_{\rm in} \propto M^{-1/4}$.
According to recent studies 
(e.g., \cite{ULX_asca2,slimdisk3,NGC1313_Suzaku,M81X9}),
the high temperature of ULXs can be explained 
if they harbor ``slim disks'' \citep{slimdisk1,slimdisk2,slimdisk3}, 
in which an advection-dominated accretion flow
and/or photon trapping (e.g., \cite{photon-trapping})
become important, due to a high accretion rate. 

Among accreting Galactic black holes that are accompanied by 
standard accretion disks,
some objects exhibit rather high temperatures, 
such as GRO J1655-40 and GRS 1915+105. 
\citet{kerrBH} proposed a spinning (i.e., Kerr) black hole scenario,
for such objects; 
the last stable orbit of a prograde accretion disk 
around a Kerr black hole can get closer to the black hole.
Accordingly $T_{\rm in}$ can potentially be higher,
compared to that of a non-rotating black hole,
although various relativistic effects should be taken into account. 
The Kerr black hole scenario was subsequently applied 
to ULXs by several authors (e.g., \cite{NGC4565,ULX_asca1}). 
Moreover, recent X-ray observations,
including those with Suzaku in particular,  
have revealed evidence of black hole rotation 
in the Seyfert I galaxy MCG$-$6$-$30$-$15 
(e.g., \cite{MCG6-30-15_Suzaku}).  

NGC 4945 is a nearby spiral galaxy with an edge-on configuration 
($\sim 80^\circ$; \cite{NGC4945_incl}).
It is regarded as a member galaxy of the Centaurus cluster,
and its distance is precisely estimated 
to be $3.7$ Mpc \citep{distance}. 
This galaxy hosts a Seyfert II nucleus 
which is one of the brightest known active galactic nuclei in the universe 
above 20 keV \citep{NGC4945_nucl_1,NGC4945_sax,NGC4945_nucl_2}.
While the nucleus is directly invisible
due to heavy obscuration below 10 keV \citep{NGC4945_ginga},
several bright X-ray point sources
are detected within the galaxy \citep{NGC4945_sax,NGC4945_xmm},
some of which are classified as ULXs \citep{ULX_xmm}.

We observed NGC 4945 twice with Suzaku \citep{Suzaku} 
separated by 5 months, 
and detected a bright transient ULX in its south-west arm region,
during the second observation in January 2006.
The observed spectral behaviour of the source indicates 
that it harbors a standard accretion disk 
rather than a slim disk, 
but simply invoking a Schwarzschild black hole leads to 
an implausible conclusion that a standard disk is shining 
at a significantly super-Eddington luminosity. 
Therefore, we have adopted the spinning black hole scenario,
and have confirmed that the observed properties can be consistently 
explained as sub-Eddington emission 
from a rapidly rotating black hole, 
with a mass of $20$ -- $130$ $M_\odot$. 

\section{Observation and Data Reduction}
\label{sec:obs}
The present observation of NGC 4945 
with the fifth Japanese X-ray observatory Suzaku \citep{Suzaku} 
were performed twice,
during the scientific working group (SWG) phase;
the first was conducted on 2005 August 22 -- 23 for about $27$ ks,
and the second one on 2006 January 15 -- 17 for about $100$ ks.
The X-ray Imaging Spectrometer (XIS; \cite{XISpaper})
and the Hard X-ray Detector (HXD; \cite{HXDdesign,HXDperform})
onboard Suzaku were operated
in the normal clocking mode with no window option
and in the normal mode, respectively.
The nucleus of NGC 4945 was placed
at the XIS-nominal and HXD-nominal positions \citep{XRTpaper}
in the first and second observations, respectively.

In the present paper, we concentrate on the XIS data;
the HXD data on the nucleus are presented 
separately by \citet{NGC4945_HXD}.
We analyzed the data with Revision 0.7 processing 
(an internal release for the SWG; \cite{NEP}), 
with the HEADAS 6.0.6 software package.
Following standard data processing,
we filtered the data by adopting the criteria 
that the spacecraft is outside the south Atlantic anomaly,
the geometric cut off rigidity is higher than $6$ GV, 
the source elevation above the rim of bright and night Earth is 
higher than $20^\circ$ and $5^\circ$ respectively, 
and the XIS data are free from telemetry saturation.
As a result, about 23 and 80 ksec of good exposures have remained
from the first and the second observations, respectively.
In the scientific analysis below,
we utilize only the events with a grade of 0, 2, 3, 4 or 6.

Because we noticed some systematic uncertainties
in the position determination in these observations,
we corrected the Suzaku attitude solution,
utilizing the archival XMM-Newton data (ObsId: 0204870101).
By referring to the positions of X-ray sources
detected with XMM-Newton (crosses in Figure \ref{fig:image})
in a region outside NGC 4945, but within the XIS field of view,
we shifted the coordinates of the XIS data by about $15''$ and $45''$, 
for the first and second observations, respectively.
These angular offsets are within nominal attitude tolerance 
of the Revision 0.7 of Suzaku data (about $1' $; \cite{XRTpaper}).

\section{X-ray Image}
\label{sec:image}
\begin{figure*}[h]
\begin{center}
\FigureFile(80mm,80mm){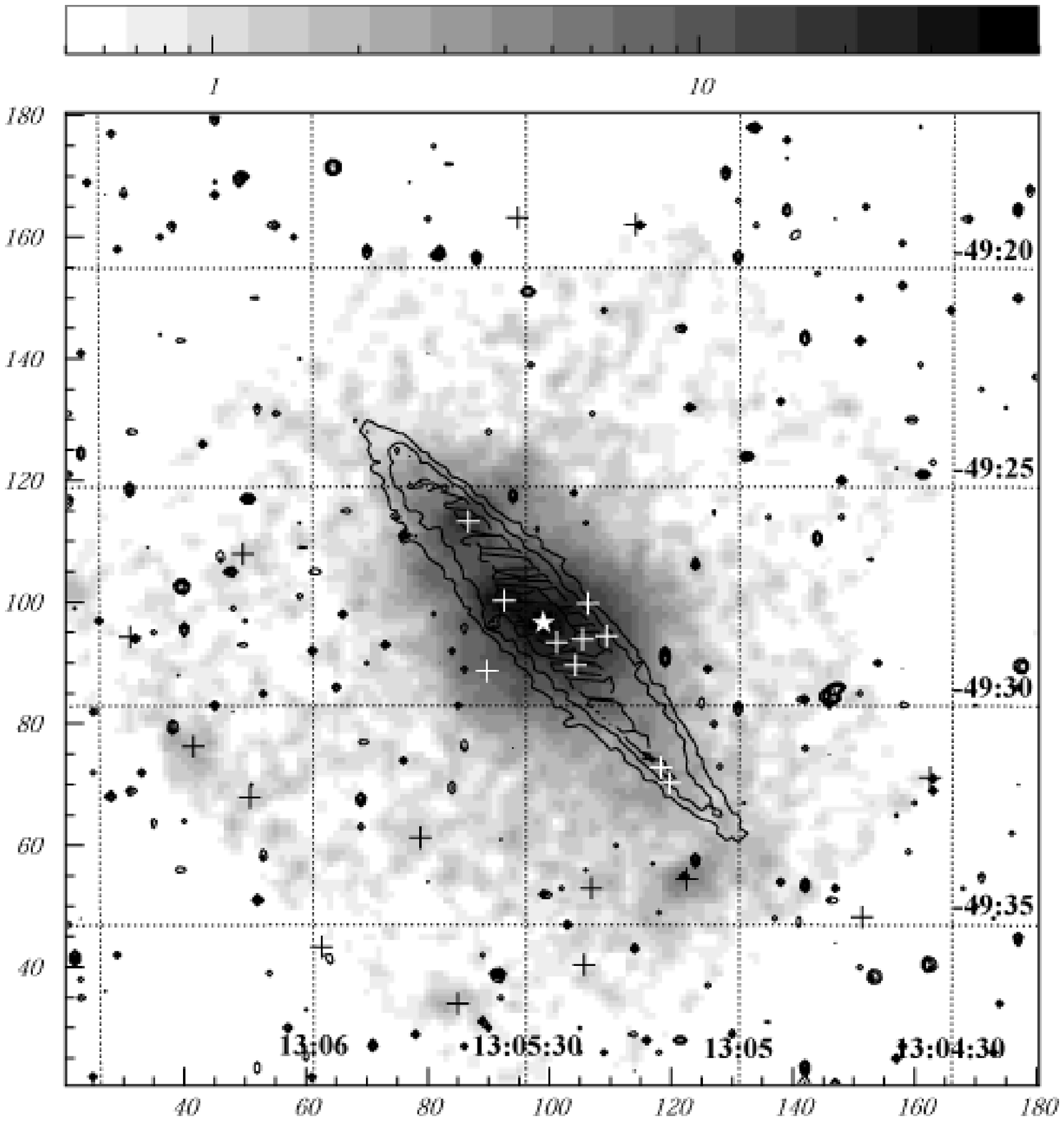}
\FigureFile(80mm,80mm){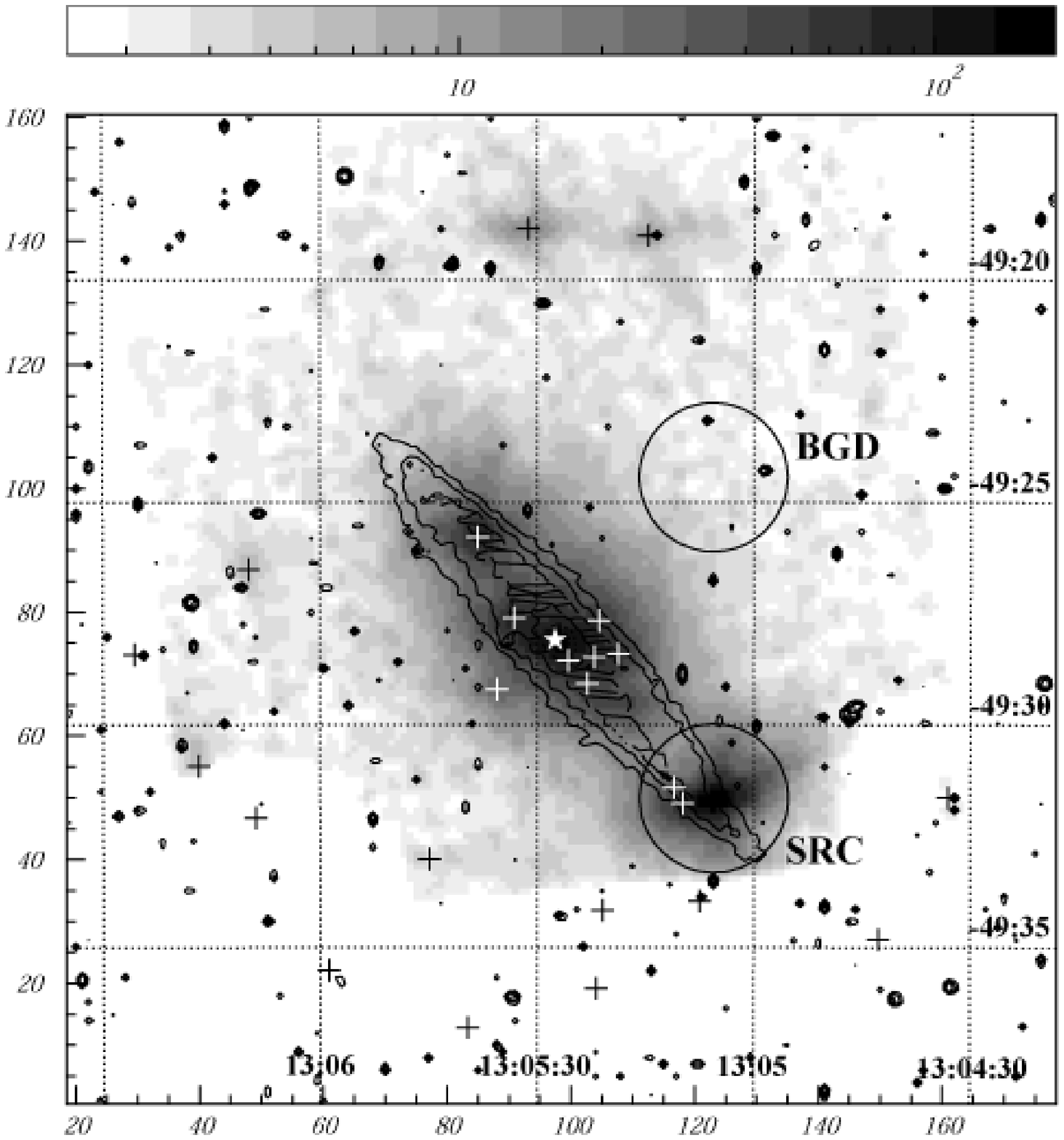}
\end{center}
\caption{
XIS FI images of NGC 4945 in 0.5 -- 10 keV,
obtained in the first (left) and second (right) observations.
The images are smoothed with a two-dimensional
Gaussian of $\sigma = 16$'' radius.
The images include background, and are not corrected for vignetting. 
Regions corresponding to the calibration radio active sources
at the XIS corners are removed.
The overlaid contours show the K-band infrared image \citep{IRimage}.
The position of the NGC 4945 nucleus is indicated by the filled star,
and X-ray sources detected with XMM-Newton by black or white crosses.
The two circles in the right panel, denoted as SRC and BGD,
show the integration regions for the source and background signals,
respectively,
employed to study Suzaku J1305-4931. 
}
\label{fig:image}
\end{figure*}

Figure \ref{fig:image} shows 0.5 -- 10 keV X-ray images of NGC 4945,
obtained with the XIS in the two observations.  
Only data from the three front-illuminated (FI) CCD sensors 
(XIS0, 2, and 3; \cite{XISpaper}) are summed up. 
An infra-red (IR) image of NGC 4945 \citep{IRimage} 
is superposed with contours in both observations. 
In addition to some point sources, we detected diffuse X-ray emission, 
elongated along the IR image of the galaxy. 
The X-ray peak of NGC 4945 coincides with the nucleus of NGC 4945,
within $8''$. 

Although the two Suzaku images are similar, 
a clear difference is a bright transient X-ray source 
which has emerged in the second observation 
in the south-west arm of NGC 4945. 
After the attitude correction,
the X-ray peak position of the transient source was determined to be
$({\rm \alpha, \delta}) =
( \timeform{13h05m05.5s}, \timeform{-49D31'39''} )$
in J2000 coordinates, with an accuracy of about \timeform{10''}.
The source is found to have no counterpart among the XMM-Newton sources 
shown in figure \ref{fig:image} (crosses),
of or recent X-ray catalogs \citep{cxo_src,xmm_src}. 
With ROSAT, \citet{NGC4945_X1} reported the detection 
of an off-center X-ray transient source, NGC 4945 X-1. 
However, the position of NGC 4945 X-1,  
$({\rm \alpha, \delta}) = 
( \timeform{13h05m22.6s}, \timeform{-49D29'12''} )$ 
with an error of \timeform{15''}, 
is clearly different from that of our Suzaku source. 
The two transients are thus considered to be distinct. 
Therefore, we regarded this Suzaku source as a new X-ray source, 
and named it Suzaku J1305-4931.

In 2005 February 8, a supernova, SN 2005aj,  
was optically discovered \citep{sn2005af} 
in the south-west of NGC 4945 near Suzaku J1305-4931. 
However SN 2005af, 
$({\rm \alpha, \delta}) =
( \timeform{13h04m44.06s}, \timeform{-49D33'59.8''} )$,  
was outside the XIS field of view in both observations.
Therefore, Suzaku J1305-4931 cannot 
be the X-ray counterpart of this supernova, either. 
In the following,  
we assume Suzaku J1305-4931
to be physically associated with NGC 4945. 

\section{The X-ray properties of Suzaku J1305-4931.  }
\label{sec:src}
\subsection{Set up for analysis}
\label{sec:src_setup}
In order to investigate the properties of Suzaku J1305-4931, 
we integrated source and background X-ray signals
within circular regions, which are denoted as {\bf SRC} and {\bf BGD}
in the right panel of figure \ref{fig:image}, respectively.
Both regions have a radius of \timeform{1.67'}, or
1.79 kpc in physical size at the distance of 3.7 Mpc.
It is difficult to adopt a larger integration circle,
because Suzaku J1305-4931 is detected 
near the edge of the XIS field of view.
The BGD region has the same separation from the NGC 4945 nucleus 
as the SRC region. 

In analyzing the XIS spectra, 
we adopted a response matrix function 
{\tt ae\_xi[0,2,3]\_20060213.rmf} for the FI CCD sensors, 
and {\tt ae\_xi1\_20060213c.rmf} for the backside-illuminated (BI) sensor
(XIS1; \cite{XISpaper}),
all of which are appropriate for the revision 0.7 data.
We discarded the 1.82 -- 1.84 keV region, 
where the XIS response is not yet fully calibrated 
due to the presence of Si-K edge. 

Due to vignetting effects of the X-ray telescope feeding the XIS,
the effective area for the SRC region in the second observation
is $\sim 40$ \% smaller than that for the XIS or HXD nominal position. 
In addition, the SRC region (\timeform{1.67'} in radius)
is significantly smaller then that of typical integration regions 
for a point source (e.g. $\sim \timeform{3'}$),
so that the effective area is further reduced.  
Therefore, we calculated the 
auxiliary response files for both observations
using a Monte-Carlo simulator {\bf xissimarfgen} \citep{xissimarf},
as of 2006 May 26 (an internal release for the SWG).
Soft X-ray absorption due to the contaminants
on the optical blocking filters 
over the XIS CCD cameras \citep{XISpaper}
were taken into account in the simulation.

\subsection{X-ray lightcurves}
\label{sec:lc}
Figure 2 shows XIS FI lightcurves of Suzaku J1305-4931 
in the soft (0.5 -- 2 keV) and hard (2 -- 10 keV) energy bands, 
accumulated over the SRC region.
The background has been subtracted utilizing the BGD region. 
After summing over the three XIS FIs, 
we measured the time-averaged net count rate of the source
in the soft and hard bands to be  
$(8.3\pm0.1)\times10^{-2}$ cts s$^{-1}$ and
$(8.7\pm0.1)\times10^{-2}$ cts s$^{-1}$, respectively.
The source thus varied by a factor of 2 on a time scale of about 
half a day in both energy bands,
with the hard band exhibiting a larger amplitude. 
The bottom panel of figure \ref{fig:ulx_lc} 
shows the hardness ratio,
simply calculated by dividing the hard band count rate by the soft band one.
It suggests that the hardness of the source changed with its brightness. 
We further examine the spectral variation in \S \ref{sec:variability}. 

\subsection{X-ray emission from the host galaxy.}
\label{sec:hostgal}
Before analyzing the X-ray spectrum of Suzaku J1305-4931, 
we briefly estimate the contribution from the NGC 4945 galaxy itself
to the SRC region.
For this purpose, we extracted the XIS spectra 
from the first observation,
and show them in figure \ref{fig:spec_1st_obs}.
The data were binned into energy intervals each containing 
at least 30 events.

We fitted the XIS spectra with a two component model,
consisting of a soft thermal emission
(a MEKAL model; \cite{MEKAL,MEKAL2,MEKAL3,MEKAL4}) and a PL model.
The former and latter are meant to represent
emission from thermal plasmas associated with the galaxy,
and the integrated emission from unresolved point sources 
within the galaxy  
(including two faint XMM-Newton sources; figure \ref{fig:image}), 
respectively.
Both components were subjected to the Galactic absorption 
toward NGC 4945, with a hydrogen column density of 
$N_{\rm H} = 1.6 \times 10^{21}$ cm$^{-2}$ \citep{NH}. 
As summarised in Table \ref{table:spec_1st_obs},
we obtained an acceptable fit with reasonable parameters 
(e.g., \cite{NGC4945_sax}).

\subsection{Time-averaged X-ray spectrum}
\label{sec:average}
Figure \ref{fig:spec_ulx} shows the XIS spectra of Suzaku J1305-4931,
obtained by accumulating events from the 2006 data over the SRC region,
and then subtracting background spectra from the BGD region. 
The FI and BI spectra were binned into energy intervals 
with at least 100 and 50 events, respectively. 
While the X-ray signals are significantly detected in 0.5 -- 10 keV,
we discarded the data below 0.75 keV in the spectral fitting,
due to a severe contribution by the emission from the host galaxy
(figure \ref{fig:spec_1st_obs}). 
The spectrum appears relatively featureless,
without any emission nor absorption lines.
We fitted the spectrum first with a PL model, 
and then with an MCD model; 
both models are widely adopted 
in studying X-ray spectra of Galactic black hole binaries and ULXs.
In the fitting, we took into account the X-ray emission 
from the host galaxy by the best-fit two-component model 
for the X-ray spectrum of the SRC region
in the first observation (Table \ref{table:spec_1st_obs}).
Since we noticed a calibration uncertainty in the flux determination
between XIS FI and BI, 
we allowed in the fittings the relative model normalization 
to deffer between FI and BI. 
As a result, XIS BI was found to yield a flux lower
by about 15 \% than XIS FI. 
Throughout the present paper,  we refer to the XIS FI flux.

The PL model gave a  best-fit photon index 
of $\Gamma = 1.89_{-0.05}^{+0.04}$, 
with a moderate fit goodness ($\chi^2 / {\rm d.o.f.} = 278.8/233$).
However, as shown in the second panel of figure \ref{fig:spec_ulx},
the fit residuals (particularly above 2 keV) imply that 
the observed spectra have a somewhat more convex shape 
than a single power law. 
This motivated us to attempt an MCD model fit. 

The MCD model gave a slightly better fit 
(the third panel of figure \ref{fig:spec_ulx}, 
$\chi^2 / {\rm d.o.f.} = 273.8/233$) than the PL one,
although the MCD model appears to be slightly too convex. 
The absorption column density, 
$N_{\rm H} = 2.17_{-0.26}^{+0.27} \times 10^{21}$ cm$^{-2}$,
is  larger by about 30 \%, 
than the Galactic value toward NGC 4945. 
The innermost temperature and innermost disk radius were 
measured to be $T_{\rm in} = 1.69_{-0.05}^{+0.06}$ keV and
$R_{\rm in} = 77.0_{-4.7}^{+5.0}~\alpha^{1/2}$ km, respectively, 
where $\alpha = (\cos 60^\circ /\cos i)$
with $i$ being the inclination of the accretion disk 
to our line of sight.
Here, we converted the apparent radius $r_{\rm in}$ to the true radius
as $R_{\rm in} = \kappa^2 \xi r_{\rm in}$, 
by adopting two correction factors (\cite{ULX_asca1} for summary); 
a spectral hardening factor of $\kappa \sim  1.7$ \citep{kappa},
which represents the ratio of the color temperature
to the effective temperature,
and a correction factor for the inner boundary condition, $\xi = 0.412$
\citep{xi}.
The bolometric luminosity of the accretion disk was calculated
to be $L_{\rm bol} = 4.4 \times 10^{39} \alpha $ erg s$^{-1}$,
using the equation
$L_{\rm bol} = 4 \pi (R_{\rm in}/\xi)^2 \sigma (T_{\rm in}/\kappa)^4$,
where $\sigma$ is the Stefan-Boltzmann constant.
This value of $L_{\rm bol}$ classifies Suzaku J1305-4931~
into the ULX category.
Figure \ref{fig:sed_ulx} shows the spectral energy distribution 
of Suzaku J1305-4931, together with the best-fit MCD model,
in the $\nu F_\nu$ form.  

We also fitted the observed spectrum with the MCD + PL model.
As shown in figure \ref{fig:spec_ulx_2comp} and Table \ref{table:spec_ulx}, 
the fit was significantly improved 
($\chi^2 / {\rm d.o.f.} = 235.2/233$) by this composite model.
The PL component dominates the MCD one above $\sim 5$ keV, 
and compensates the spectral curvature more concave 
than the single MCD model. 
The MCD temperature, $T_{\rm in} = 1.25_{-0.11}^{+0.24}$ keV,   
becomes slightly lower than that of the single MCD fit. 
However, the PL component turned out to be rather hard, 
$\Gamma = 1.23_{-1.47}^{+0.51}$, compared with the typical value of ULXs 
with MCD+PL-type X-ray spectra 
($\Gamma \gtrsim 2.0$; e.g, \cite{MCD_PL, MCD+PL_2}).
Therefore, we regard this composite model as physically 
of little meaning, and do not discuss it in any further. 
Incidentally, this MCD + PL model fit did not converge to a ``cool''  disk 
(e.g., $T_{\rm in} \sim 0.2$ keV) disk solution,
which was reported by some authors (e.g., \cite{cool_disk}). 
Likewise, based on the Suzaku data on NGC 1313 X-1, 
\citet{NGC1313_Suzaku} cast a doubt upon the reality of such 
``cool'' disks in ULXs.

\subsection{X-ray Spectral Variability}
\label{sec:variability}
Since figure \ref{fig:ulx_lc} 
suggests the intensity-correlated spectral variation of Suzaku J1305-4931, 
we plotted in figure \ref{fig:ulx_hardness},
the hardness ratio against the total (0.5 -- 10 keV) count rate. 
The plot shows a clear trend that 
the spectrum of the source gets harder, as it becomes brighter. 
Therefore, we divided the total exposure into 4 periods
according to the intensity of the source,
and investigated in detail the variation of the spectral parameters. 
The definition of these periods are indicated 
with dashed-lines in figure \ref{fig:ulx_lc}, 
and their effective exposures are shown 
in table \ref{table:spec_variability}.

Figure \ref{fig:spec_variability} shows the XIS spectra 
of these 4 periods.
We fitted them by the MCD model,  
because it gave a better fit to the time-averaged spectra 
than the PL model.
The absorbing column density was fixed at the best-fit value 
from the time-averaged spectra shown
in table \ref{table:spec_ulx}, 
$N_{\rm H} = 2.17 \times 10^{21}$ cm$^{-2}$. 
The model successfully described the four spectra 
with the best-fit parameters tabulated 
in table \ref{table:spec_variability}.
The source shows the highest temperature as
$T_{\rm in} = 1.77_{-0.08}^{+0.09}$ keV (i.e. the hardest spectrum)
in the brightest period
(Period 1, $L_{\rm bol} = 5.9 \times 10^{39} \alpha $ erg s$^{-1}$),
while the lowest temperature as $T_{\rm in} = 1.51_{-0.11}^{+0.13}$ keV
in the faintest period
(Period 4, $L_{\rm bol} = 3.0 \times 10^{39} \alpha $ erg s$^{-1}$).
This is consistent with the behaviour of the hardness ratio 
shown in figures \ref{fig:ulx_lc} and \ref{fig:ulx_hardness}. 

Throughout the observation, the innermost radius of the accretion disk 
has been found to stay relatively constant
at $R_{\rm in} \sim 80 \alpha^{1/2} $ km.
Therefore, we re-fitted the spectra of the four periods,
by tying $R_{\rm in}$ together, 
but allowing $T_{\rm in}$ to take independent values. 
We also adopted a common relative normalization factor between XIS FI and BI. 
A similar fit goodness has been obtained ($\chi^2 / {\rm d.o.f.} = 485.0/516$), 
in comparison with a simple sum over the four individual fits 
($\chi^2 / {\rm d.o.f.} = 484.4/510$ in total);
the other parameters remained unchanged within statistical uncertainties.
As a result, the disk radius has been obtained as 
$R_{\rm in}= 79.1_{-3.9}^{+4.0}~\alpha^{1/2}$ km,
which agrees, within errors, with that from the time-averaged spectrum.

\section{Discussion}
\label{sec:discussion}
\subsection{Summary of the results}
In the second Suzaku observation 
of the nearby Seyfert II galaxy NGC 4945,
performed on 2006 January 15 -- 17, 
we have discovered a new bright transient ULX, Suzaku J1305-4931,
within the south-west arm region.
We assume the source to be physically associated with NGC 4945. 
The time-averaged XIS spectrum of the source was approximated 
by a PL model with photon index $\sim 1.9$, 
or somewhat better by an MCD model with $T_{\rm in}\sim 1.7$ keV. 
Adopting the MCD model, the bolometric luminosity of the source 
was determined to be 
$L_{\rm bol} =  4.4 \times 10^{39} \alpha $ erg s$^{-1}$. 
These values lie within the typical range of those of well-studied ULXs
(e.g. \cite{ULX_asca1,ULX_asca2}).
During the observation,
Suzaku J1305-4931 varied by a factor of 2  
in intensity in a timescale of about a half day.

Suzaku J1305-4931 was undetectable in the first Suzaku observation,
conducted in 2005 August 22 -- 23,
and has no counterpart among the available X-ray source catalogs 
\citep{cxo_src,xmm_src}.
We fitted the XIS spectrum from the first observation, 
by adding an MCD component to the emission from the host galaxy
(Figure \ref{fig:spec_1st_obs} and Table \ref{table:spec_1st_obs}).
Then, the upper limit on the $0.5$ -- $10$ keV source flux was obtained 
to be $1.7 \times 10 ^{-14}$ erg cm$^{-2}$ s$^{-1}$ 
(at 90\% confidence level).
Therefore, the source is inferred to have brightened
by more than two orders of magnitude in about 5 mounts. 

The MCD model fit gives an innermost disk radius of 
$R_{\rm in} = 77.0_{-4.7}^{+5.0}~\alpha^{1/2}$ km. 
Equating $R_{\rm in}$ to the last stable orbit
of a standard accretion disk \citep{standard_disk}
around a Schwarzschild black hole,
$3 R_{\rm s}$ where $R_{\rm s}$ is the Schwarzschild radius,
the mass of the source would become 
$M = (R_{\rm in} / 8.86~{\rm km}) M_\odot
= 8.7_{-0.5}^{+0.6} \alpha^{1/2} M_\odot$.
However, the Eddington limit for this mass is calculated as 
$L_{\rm Edd} = 1.3 \times 10^{39} \alpha^{1/2} $ erg s$^{-1}$. 
Therefore, the source would be concluded to be shining
at the Eddington ratio of 
$\eta = L_{\rm bol} / L_{\rm Edd} = 3.4 \alpha^{1/2}$.
Thus, the source appears to show the ``super-Eddington'' problem.

\subsection{Slim disk or standard disk ?}
The ``super-Eddington'' problem is commonly found 
among ULXs showing an MCD type X-ray spectrum. 
Usually, the problem is solved by invoking the slim disk model 
(e.g, \cite{ULX_asca2,slimdisk3,M81X9}),  
which is thought to be realized at a high accretion rate
\citep{slimdisk1,slimdisk2,slimdisk3}. 
In a slim disk, an advection-dominated accretion flow,
and/or photon trapping (e.g., \cite{photon-trapping}) become important.
As a result, 
the X-rays are expected to be radiated 
from inside the last stable orbit of the accretion disk as well, 
and the local temperature profile is predicted to become flatter 
than the standard disk \citep{slimdisk3}.

In order to examine the present data for the possible presence 
of a slim disk,  
we analyzed the time-averaged XIS spectrum with an accretion disk model,
in which the local disk temperature is proportional to $r^{-p}$ 
with $p$ being a positive free parameter, where $r$ is a local radius
\citep{pfree_disk,pfree_disk_2,XTEJ1550}.
While the MCD model assumes the fixed value of $ p = 3/4$,
integrated emission from a slim disk is thought to be approximated 
by smaller value of $p$, down to $p=1/2$ \citep{slimdisk3}.
Here, we denote this model the ``variable-$p$'' disk model. 
A good fit was obtained by this model 
(the forth panel of figure \ref{fig:spec_ulx}; 
$\chi^2 / {\rm d.o.f.}  =243.5/232$),
together with the parameters shown in table \ref{table:spec_ulx}. 
Compared with the MCD fit, 
the disk temperature became higher,
$T_{\rm in} = 2.31_{-0.23}^{+0.45}$ keV, and 
the disk radius got smaller, 
$R_{\rm in} = 37.6_{-13.0}^{+11.5}~\alpha^{1/2}$ km.  
We obtained $p = 0.59\pm0.03$.
This value of $p$ is consistent with the fact that 
the observed X-ray spectrum is slightly more convex than the PL model,
but slightly more concave than the MCD one.
All these parameters appear to be reasonable 
for the slim disk picture. 

The observed spectral variation of Suzaku J1305-4931,
on the other hand, 
seems to contradict the slim disk interpretation. 
In Figure \ref{fig:Lbol-kT},
we plot $L_{\rm bol}$ against $T_{\rm in}$ of the source,
for the 4 individual periods (see Table \ref{table:spec_variability}).
For comparison, we also compiled those of some well-studied Galactic
and Magellanic black holes,
and ULXs \citep{ULX_asca1,NGC2403,ULX_asca2,NGC4449,NGC253X21,M81X9}.
The super-Eddington problem of Suzaku J1305-4931 
is clearly visualised in the figure,
in which the data points for the source are located 
above the line of $\eta =1$.
However, 
it is important to note that Suzaku J1305-4931 shows 
a clear correlation as $L_{\rm bol} \propto T_{\rm in}^4$.
Correspondingly, the innermost disk radius stays constant 
around $R_{\rm in}= 79.1_{-3.9}^{+4.0}~\alpha^{1/2}$ km,
which is consistent with the value of the time-averaged spectrum.
Such correlations have been frequently observed
in Galactic/Magellanic black holes (e.g.,\cite{LMCX3,XTEJ1550}),
and is considered to be a signature of a standard accretion disk
\citep{standard_disk}.
On the other hand,
the MCD model fit to the numerically simulated slim disk spectra 
is reported to give a relation of $L_{\rm bol} \propto T_{\rm in}^2$
\citep{slimdisk3}, and various ULXs actually follow this relation 
(Figure \ref{fig:Lbol-kT}; \cite{ULX_asca2}). 

For a further examination,
we simulated the dependence of the hardness ratio 
on the 0.5 -- 10 keV count rate for various values of $T_{\rm in}$,
assuming either $L_{\rm bol} \propto T_{\rm in}^4$ or
$L_{\rm bol} \propto T_{\rm in}^2$.
We utilized the same response files as used in the spectral fitting,
and took into account the contribution of the emission 
from the host galaxy (table \ref{table:spec_1st_obs}). 
The model predictions are shown in figure \ref{fig:ulx_hardness}, 
in comparison with the observed data.
The standard disk ($L_{\rm bol} \propto T_{\rm in}^4$) locus 
is found to describe the observed relation 
between the hardness and count rate with $\chi^2 / {\rm d.o.f.} = 9.9 / 11$ 
(only the erros of the hardness ratio are considered),
which is better than is obtained with the slim disk 
($L_{\rm bol} \propto T_{\rm in}^2$) one $\chi^2/{\rm d.o.f.} = 12.7/11$.
Therefore, the source variation is described better by the 
$L_{\rm bol} \propto T_{\rm in}^4$ relation than by the 
$L_{\rm bol} \propto T_{\rm in}^2$ dependence.

Except for Galactic/Magellanic sources, 
a luminous source, NGC 253 X21 \citep{NGC253X21}, 
with a luminosity of $L_{\rm bol} \lesssim 10^{39}$ erg s$^{-1}$,
is reported to follow a similar $L_{\rm bol} \propto T_{\rm in}^4$
spectral behavior.
However, this correlation has not been found so far from ULXs 
with a luminosity higher than a few $10^{39}$ erg s$^{-1}$. 
Therefore, we regard Suzaku J1305-4931 
to be the most luminous black hole
that shows the $L_{\rm bol} \propto T_{\rm in}^4$ relation,  
and hence is thought to host a standard accretion disk,
although a simple standard disk cannot 
solve the super-Eddington problem.

\subsection{Kerr black hole scenario}
We have to look for some other mechanisms to explain
the high luminosity and high temperature of the source.
One of the most important remaining physical parameters
is a possible rotation of the black hole.
The rotating black-hole scenario (i.e., a Kerr black hole) 
was originally proposed by \citet{kerrBH}
to explain the high disk temperature of the Galactic jet source,
GRS 1915+105 and GRO J1655-40.
The idea was subsequently applied to
the interpretation of the X-ray properties 
of the ULXs observed with ASCA \citep{NGC4565,ULX_asca1}.
Rapid black-hole spin is expected to make 
the last stable orbit of an accretion disk around it 
come closer to the black hole, down to $0.5 R_{\rm s}$ 
for the most extreme case.
Neglecting the general relativistic effects,
the Kerr black hole assumption can enlarge the mass 
estimated by equating $R_{\rm in}$ to the last stable orbit, 
by up to 6 times.
Simultaneously, the maximum value of $T_{\rm in}$
at the Eddington accretion rate can increase by a factor of $\sqrt 6$.
Moreover, the radiation efficiency of a Kerr black hole
can be significantly higher than that of a Schwarzschild black hole
\citep{kerrBH}.
Therefore, the Kerr black hole scenario is thought to give
a possible solution to the super Eddington problem of Suzaku J1305-4931.

In order to quantitatively evaluate the possibility
of black hole rotation in Suzaku J1305-4931,
we adopted the {\bf KERRBB} model in the XSPEC package \citep{kerrbb}.
The model describes the numerically computed radiation spectrum
from a geometrically thin accretion disk around a Kerr black hole,
and adequately takes into account various general relativistic effects.
It is intended to extend the {\bf GRAD} model 
in XSPEC \citep{grad1,grad2},
which represents the general relativistic accretion disk
around a Schwarzschild black hole.
\citet{kerrbb} applied this model to three Galactic sources,
for which the mass and disk inclination are already known; 
GRO J1655-40, 4U 1543-47, and XTE J1550-564. 
They concluded that GRO J1655-40 and 4U 1543-47 have 
a high black-hole spin, 
while XTE J1550-564 hosts a Schwarzschild black hole; 
the results on GRO J1655-40 is consistent with a speculation 
by \citet{GROJ1655_spin}
based on its comparison with LMC X-3. 
Therefore, the KERRBB model is quite suitable for our purpose. 

We re-analysed the time-averaged spectrum 
of Suzaku J1305-4931~with KERRBB.
Among the nine physical parameters in the model,
we switched off the effects of limb-darkening, self-irradiation and
torque at the inner boundary of the accretion disk,
since these are relatively unimportant for our purpose \citep{kerrbb}.
We again adopted a spectral hardening factor of $\kappa = 1.7$
\citep{kappa}.
Then, the emergent spectrum should be determined by specifying
the mass $M$, 
mass accretion rate $\dot{M}$, 
disk inclination $i$, and spin parameter $a= c / (GM^2) J$,
where $c$, $G$, and $J$ are the light speed, the gravitational constant,
and angular momentum of the black hole, respectively.
However, the observations can provide
basically only two constraints, namely the flux and color
temperature (except detailed spectral shapes). 
As a result, the KERRBB solutions are degenerate, 
so that we need to specify both $a$ and $i$ 
to estimate $M$ and $\dot{M}$ from the data.
Thus, we calculated the best-fit values of $M$ and $\dot{M}$,
for some representative values of $i$ and $a $. 
When we increase $a$ with $i$ fixed, the emergent flux and    
color temperature both increases, because the last stable
orbit gets closer to the black hole. Then, the model tries to reduce
the color temperature by increasing $M$, 
while reducing the flux by lowering $\dot{M}$. 
This $a$-dependence become more enhanced    
toward higher inclination, because the longitudinal Doppler
effects of the disk rotation get stronger.
Here, we only considered the case of $a \ge 0$
(i.e., a prograding accretion disk)
and adopted the maximum value of $a = 0.998$ \citep{max_spin}. 

Acceptable fits have been obtained for a variety of $i$ and $a$
($\chi^2 / {\rm d.o.f.} = 1.07 \sim 1.18$);
the obtained values of $M$ and $\dot{M}$
are summarised in figure \ref{fig:kerrBB},
where we also plot the MCD model results, 
in order to illustrate the relativistic effects 
which are considered in the KERRBB model 
but not in the MCD approximation.
In the figure, the Eddington ratio $\eta$ is indicated with red lines.
We regard that the region below the line of $\eta = 1$ 
is physically meaningful. 
Thus, the observed X-ray spectrum of Suzaku J1305-4931~can be 
interpreted as a black hole with $a \sim 1$ and $i = 50$ --$80^\circ$,
radiating at $\eta \lesssim 1$.
Especially, a combination of $a=0.998$ and $i=60^\circ$ 
gives the most successful description of the observed spectrum
($\chi^2 / {\rm d.o.f.} = 249.3/232$), 
yielding $M = 36 M_\odot$ and $\dot{M} = 2.0 \times 10^{19}$ g s$^{-1}$.
The fit residuals for these parameters 
are shown in the bottom panel of figure \ref{fig:spec_ulx}.
Therefore, we conclude that a nearly extreme Kerr black hole 
with a mass of $M = 20$ -- $130 M_\odot$
is one of the most promising candidates 
for the nature of Suzaku J1305-4931.

\begin{figure}[h]
\begin{center}
\FigureFile(80mm,80mm){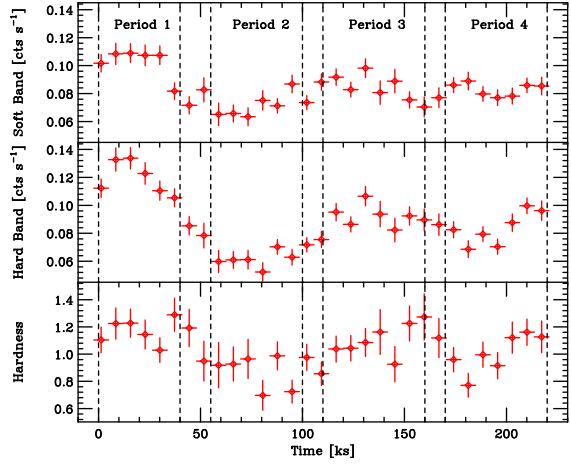}
\end{center}
\caption{
Background-subtracted XIS FI lightcurves of Suzaku J1305-4931,
binned into 2 hours (7200 sec).
The top and middle panels show the soft (0.5 -- 2 keV)
and hard (2 -- 10 keV) band lightcurves, respectively.
The hardness of the source,
defined as the ratio of the hard band count rate to the soft band one,
is shown in the bottom panel.
The definition of 4 periods,
adopted in the study of spectral variations,
is indicated with dashed lines.}
\label{fig:ulx_lc}
\end{figure}

\begin{figure}[h]
\begin{center}
\FigureFile(80mm,80mm){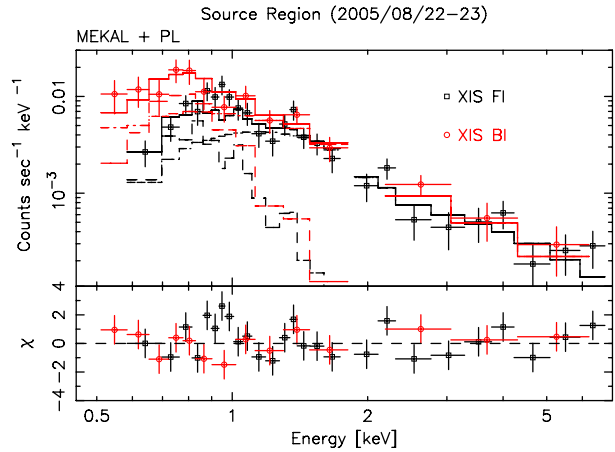}
\end{center}
\caption{
XIS spectra of the SRC region obtained in the first observation.
The FI and BI data are shown with the black and red data points, 
respectively.
The best-fit model, consisting of a MEKAL and a PL component,
are shown with histograms. }
\label{fig:spec_1st_obs}
\end{figure}

Among Galactic/Magellanic black holes, 
GRS 1915+105 is the only source 
exhibiting both a comparable high Eddington ratio 
($\eta = 0.2$ -- $3$, \cite{GRS1915_3}),
and a high spin parameter estimated ($a \sim 0.7$, \cite{GRS1915_spin}).
One of the most prominent characteristics of GRS 1915+105.
especially in the super-Eddington phase (\cite{GRS1915_3}), 
is its rapid variability with a timescale 
from $\sim$ an hour to even down to sub-seconds, 
presumably due to a disk instability (e.g., \cite{GRS1915_1, GRS1915_2}). 
Although it is interesting to ask whether such a variability is 
present in Suzaku J1305-4931 as well, 
we can only say, due to the limited statistics (figure \ref{fig:ulx_lc}),
that any short-term intensity variation is less than by 10 \% and 15 \%, 
on representative time scale of $1$ hour and $10$ minutes, respectively.

We are grateful to all the members of the Suzaku team,
for the successful operation and calibration.
We also thank Prof. R. Narayan for kindly giving us information
on the KERRBB model.
The constructive comments from the anonymous referee 
greatly helped us to improve the quality of the paper. 
N. I. is supported by the Special Postdoctoral Researchers Program 
in RIKEN.
P. G. is a Fellow of the Japan Society for Promotion of Science (JSPS).
We have made extensive use of the NASA/IPAC Extra galactic Database
(NED; the Jet Propulsion Laboratory, California Institute
of Technology, the National Aeronautics and Space Administration).


\clearpage
\begin{table}[htbp]
\caption{The best-fit spectral parameters 
of the SRC region for the first observation.}
\label{table:spec_1st_obs}. 
\begin{center}
\begin{tabular}{ll}
\hline\hline 
Parameters                &  Value \\
\hline       
$N_{\rm H}$ ($10^{21}$ cm$^{-2}$) \footnotemark[$*$] 
                          & $1.6\times10^{21}$            \\
$kT$ (keV)                & $0.34_{-0.06}^{+0.26}$        \\
$F_{\rm th}$ ($10^{-13}$ erg s$^{-1}$ cm$^{-2}$) 
\footnotemark[$\dagger$]  & $0.33$  \\
$\Gamma_{\rm PL}$         & $2.16_{-0.28}^{+0.17}$        \\
$F_{\rm PL}$ ($10^{-13}$ erg s$^{-1}$ cm$^{-2}$) 
\footnotemark[$\ddagger$] & $1.6$        \\
$\chi^2/{\rm d.o.f.}$      & $43.1/36$     \\
\hline       
\multicolumn{2}{@{}l@{}}{\hbox to 0pt{\parbox{60mm}{\footnotesize
\footnotemark[$*$]       Fixed at the Galactic value. 
\par\noindent
\footnotemark[$\dagger$]  
Absorption-inclusive 0.5 -- 10 keV model flux of the thermal component.
\par\noindent
\footnotemark[$\ddagger$] 
Absorption-inclusive 0.5 -- 10 keV model flux of the PL component. 
}\hss}}
\end{tabular}
\end{center}
\end{table}

\begin{longtable}{lllll}
\caption{The best-fit parameters 
for the time-averaged spectrum of Suzaku J1305-4931.}
\label{table:spec_ulx}
\hline\hline 
Parameters  
       &  PL            & MCD                     
       & MCD + PL       &variable-$p$ \\ 
\hline 
\endfirsthead
\hline\hline
Parameters                            &  PL                     
            & MCD                     & MCD + PL       
            & variable-$p$               \\ 
\hline
\endhead
\hline
\endfoot
\hline
\multicolumn{5}{@{}l@{}}{\hbox to 0pt{\parbox{100mm}{\footnotesize
\footnotemark[$*$] 
Innermost disk radius, assuming an inclination of \timeform{60D}
\par\noindent
\footnotemark[$\dagger$] 
$0.5$ -- $10$ keV source flux in units of 
$10^{-12}$ erg cm$^{-2}$ s$^{-1}$, without correction for absorption.  
\par\noindent
\footnotemark[$\ddagger$] 
flux of PL / MCD components
}\hss}}
\endlastfoot
$N_{\rm H}$ ($10^{21}$ cm$^{-2}$)      & $5.82_{-0.41}^{+0.43}$  
             & $2.17_{-0.26}^{+0.27} $ & $3.14_{-0.63}^{+1.06}$  
             & $4.12_{-0.53}^{+0.64} $                             \\
$T_{\rm in}$ (keV)                     & --                      
             & $1.69_{-0.05}^{+0.06} $ & $1.25_{-0.11}^{+0.24}$  
             & $2.31_{-0.23}^{+0.45} $                             \\
$R_{\rm in} $ (km) \footnotemark[$*$]  & --                      
             & $77.0_{-4.7}^{+5.0}$    & $107.6_{-39.0}^{+23.1}$  
             & $37.6_{-13.0}^{+11.5}$                              \\  
$\Gamma$ or $p$                        & $1.89_{-0.05}^{+0.04}$  
             & --                      & $1.23_{-1.47}^{+0.51}$  
             & $0.59\pm0.03 $                                      \\
$F_{\rm X}$ \footnotemark[$\dagger$]   & $2.4$                   
             & $2.2$           & $1.2 / 1.2$ \footnotemark[$\ddagger$]  
             & $2.2$                                               \\
$\chi^2/{\rm d.o.f.}$                   & $278.7/233 $            
             & $273.9/233$             & $235.2/231$  
             & $243.5/232 $                                        \\
\end{longtable}

\begin{longtable}{lllll}
\caption{The spectral variation of Suzaku J1305-4931.}
\label{table:spec_variability}
\hline\hline 
Parameters                                       
      & Period 1 & Period 2 & Period 3 & Period 4 \\
\hline 
\endfirsthead
\hline\hline
Parameters                                       
      & Period 1 & Period 2 & Period 3 & Period 4 \\
\hline
\endhead\hline
\endfoot\hline 
\multicolumn{5}{@{}l@{}}{\hbox to 0pt{\parbox{100mm}{\footnotesize
\footnotemark[$*$ ] 
  Innermost disk radius, assuming an inclination of \timeform{60D}
\par\noindent
\footnotemark[$\dagger$] 
  Absorption-inclusive $0.5$ -- $10$ keV source flux, 
  in units of $10^{-12}$ erg cm$^{-2}$ s$^{-1}$.
\par\noindent
\footnotemark[$\ddagger$] 
  Bolometric luminosity in units of $10^{39}$ erg s$^{-1}$, 
  assuming an inclination of \timeform{60D} 
}\hss}}
\endlastfoot
Exposure (ksec)                         
   & $15.8$                    & $13.2$                   
   & $17.1$                    & $19.9$ \\
$N_{\rm H}$ ($10^{21}$ cm$^{-2}$)       
   & \multicolumn{4}{c}{$2.17$ (fix)}            \\
$T_{\rm in}$ (keV)             
   & $1.77_{-0.08}^{+0.09}$    & $1.51_{-0.11}^{+0.12}$    
   & $1.68_{-0.09}^{+0.10}$    & $1.65_{-0.08}^{+0.09}$\\
$R_{\rm in} $ (km) \footnotemark[$*$]   
   & $80.9_{-6.8}^{+7.2}$      & $80.2_{-10.6}^{+11.8}$   
   & $77.9_{-7.3}^{+7.8}$      & $77.7_{-6.9}^{+7.5}$ \\
$F_{\rm X}$ \footnotemark[$\dagger$]    
   & $2.9$                     & $1.5$                    
   & $2.2$                     & $2.0$                \\
$L_{\rm bol}$ \footnotemark[$\ddagger$] 
   & $5.9$                     & $3.0$                    
   & $4.4$                     & $4.1$                \\
$\chi^2/{\rm d.o.f.}$                    
   & $135.3 / 153$             & $85.7/79$                
   & $119.8 / 132 $            & $143.6 / 146 $
\end{longtable}


\begin{figure}[h]
  \begin{center}
    \FigureFile(80mm,80mm){figure4.ps}
  \end{center}
  \caption{
XIS spectra of Suzaku J1305-4931.
The black and red points indicate the FI and BI data, respectively.
The best-fit MCD model is shown with the histograms 
in the first panel,
where the contribution from the host galaxy is represented 
by two fixed model components. 
The second, third, forth and fifth panels display the residuals 
from the PL, MCD, variable-$p$ disk, 
and KERRBB (\cite{kerrbb}; with $i=60^\circ$ and $a=0.998$) fits, 
respectively.
}
\label{fig:spec_ulx} 
\end{figure}

\begin{figure}[h]
\begin{center}
\FigureFile(80mm,80mm,angle=-90){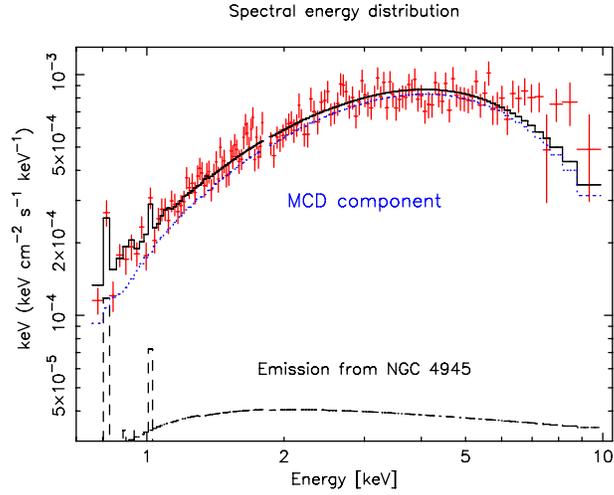}
\end{center}
\caption{Spectral energy distribution of Suzaku J1305-4931,
in the $\nu F_{\rm \nu}$ form.
Only the FI data are shown, for clarity. 
The histograms shows the best-fit MCD model,
including the emission from the host galaxy.}
\label{fig:sed_ulx} 
\end{figure}

\begin{figure}[h]
\begin{center}
\FigureFile(80mm,80mm){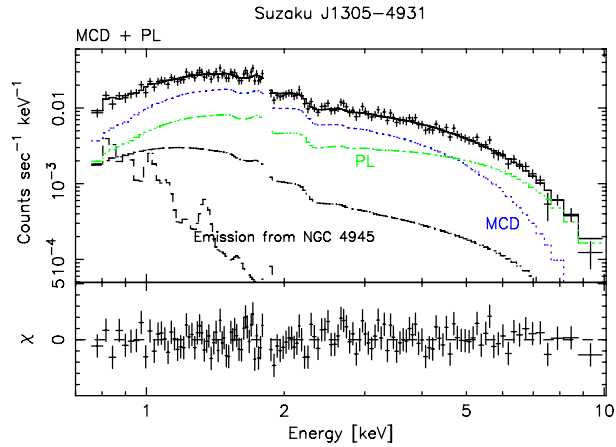}
\end{center}
\caption{The MCD + PL fit to the XIS spectra.
Although the same data as in figure \ref{fig:spec_ulx} are utilized, 
only the FI data are shown for clarity. 
The blue and green histograms indicate 
the best-fit MCD and PL component, respectively. 
}
\label{fig:spec_ulx_2comp} 
\end{figure}

\begin{figure}[h]
\begin{center}
\FigureFile(80mm,80mm){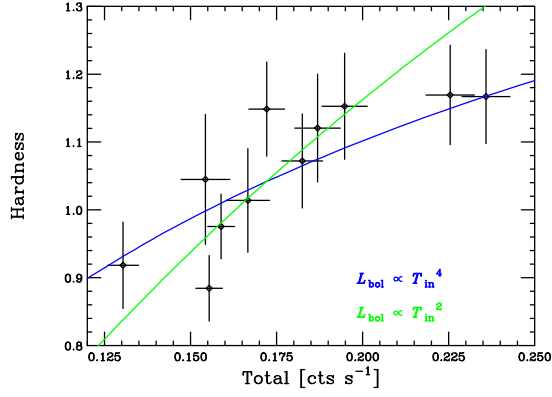}
\end{center}
\caption{Hardness ratio plotted 
against the total count rate in 0.5 -- 10 keV. 
The blue and green lines indicate the prediction 
for the case of $L_{\rm bol} \propto T_{\rm in}^4$ 
(i.e., $R_{\rm in}$ is constant) 
and $L_{\rm bol} \propto T_{\rm in}^2$, respectively.
The source variation is assumed to occur 
around a reference condition of $T_{\rm in} = 1.69$ keV and 
$L_{\rm bol} = 4.4 \times 10^{39}$ erg s$^{-1}$, 
as determined from the MCD fitting to 
the time-averaged spectrum shown in figure \ref{fig:spec_ulx}.
Each data point is an integration for $15$ ks. 
} 
\label{fig:ulx_hardness}
\end{figure}

\begin{figure*}[h]
\begin{center}
\FigureFile(60mm,60mm){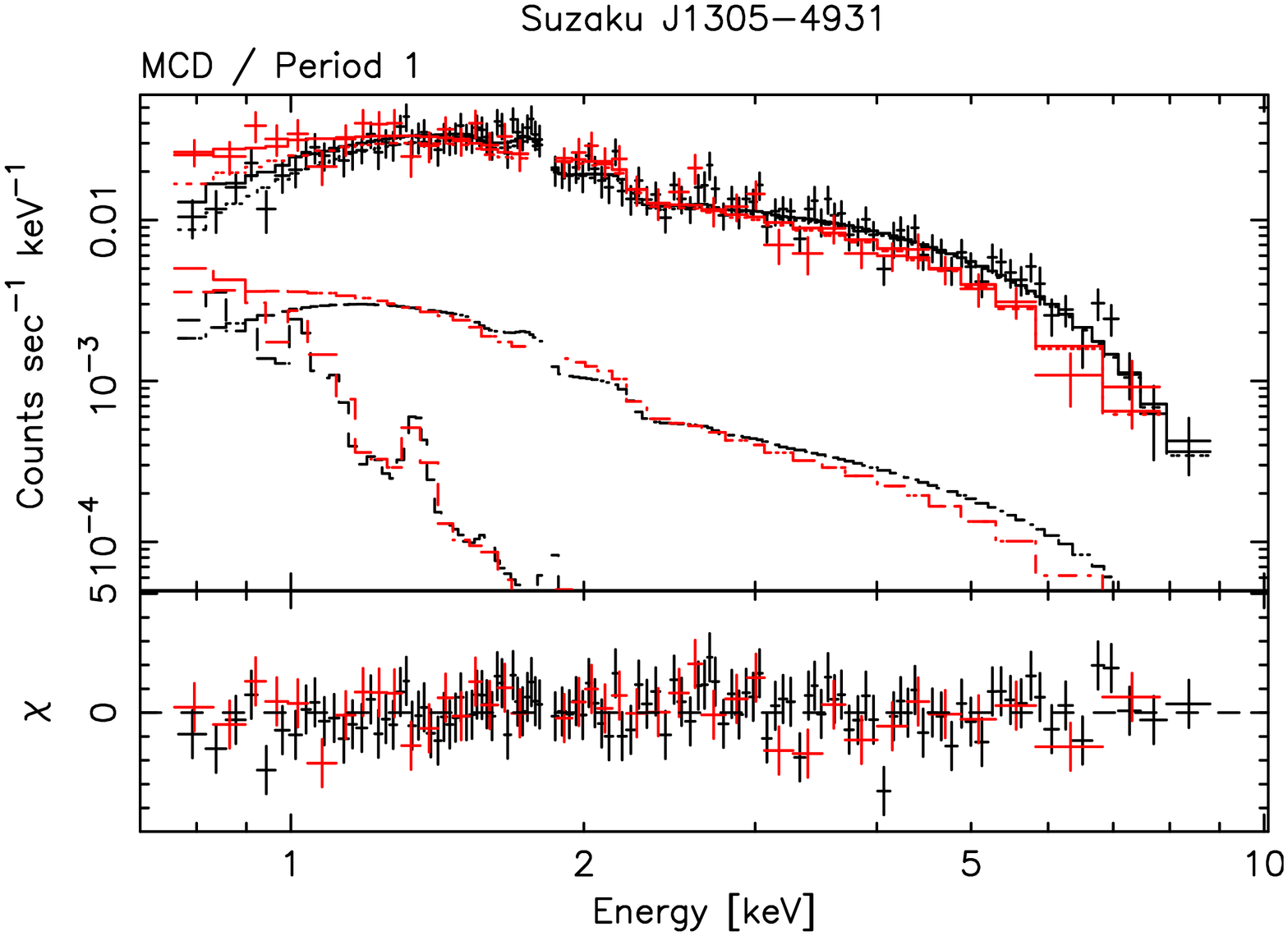}
\FigureFile(60mm,60mm){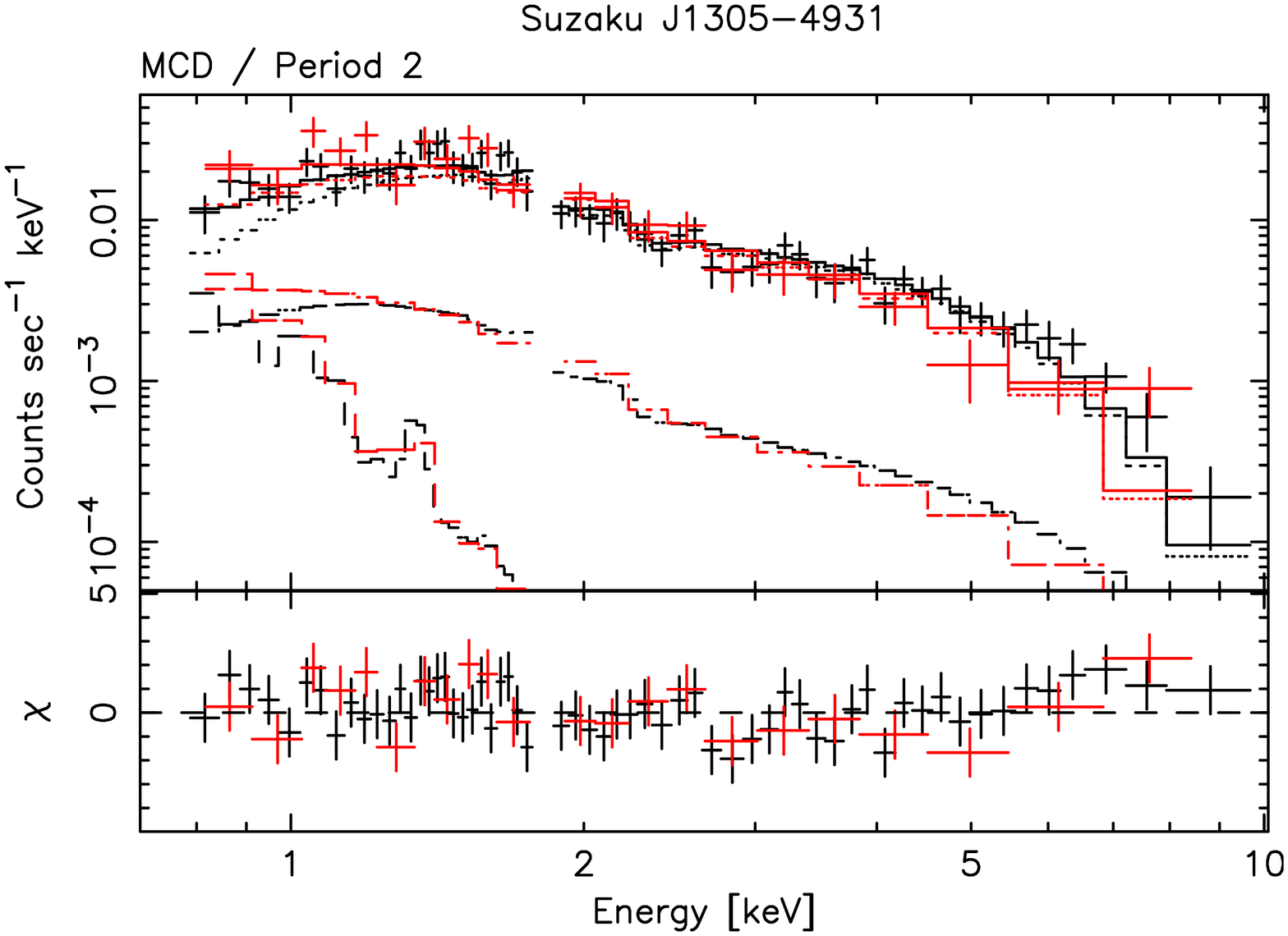}
\FigureFile(60mm,60mm){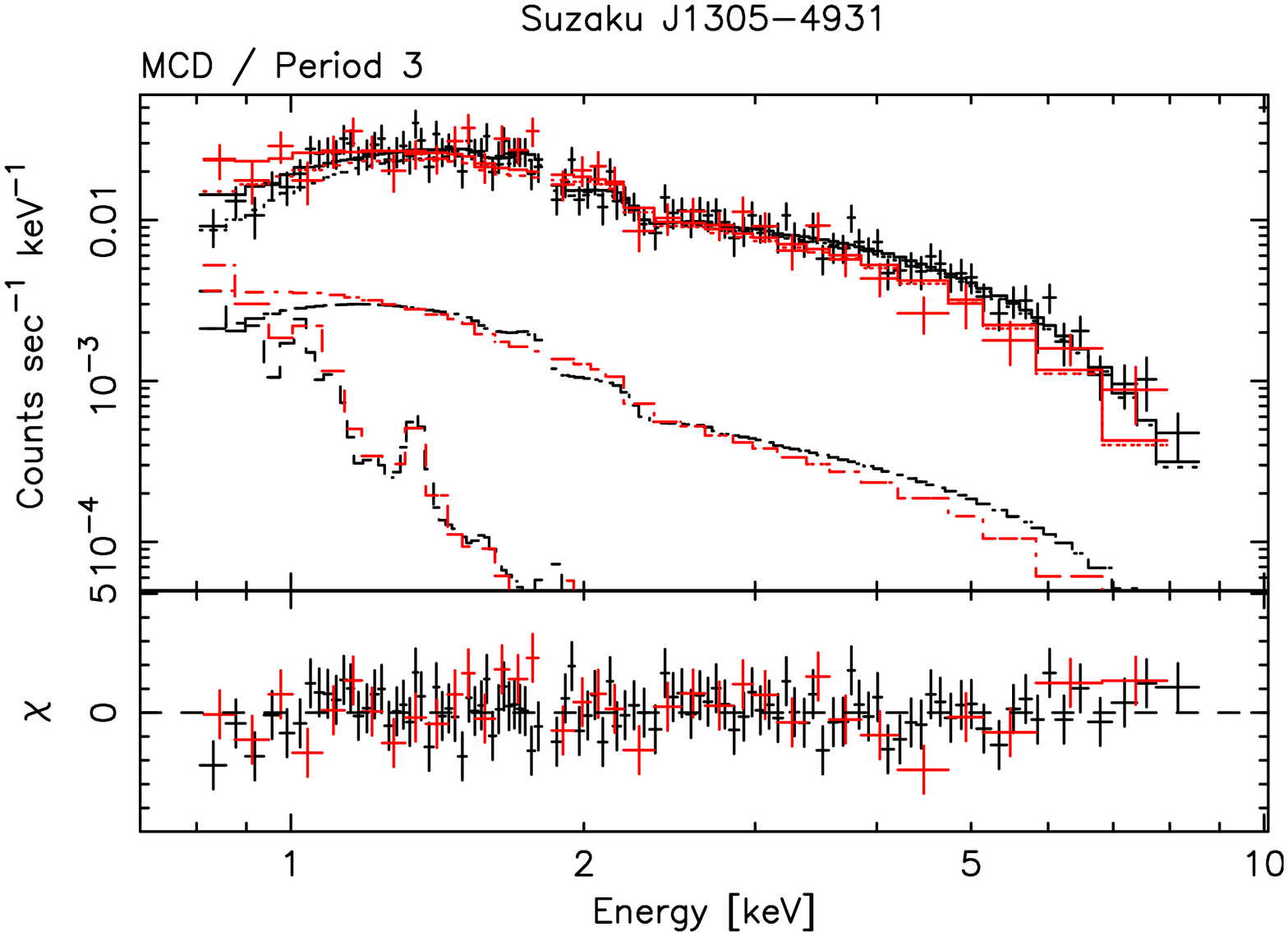}
\FigureFile(60mm,60mm){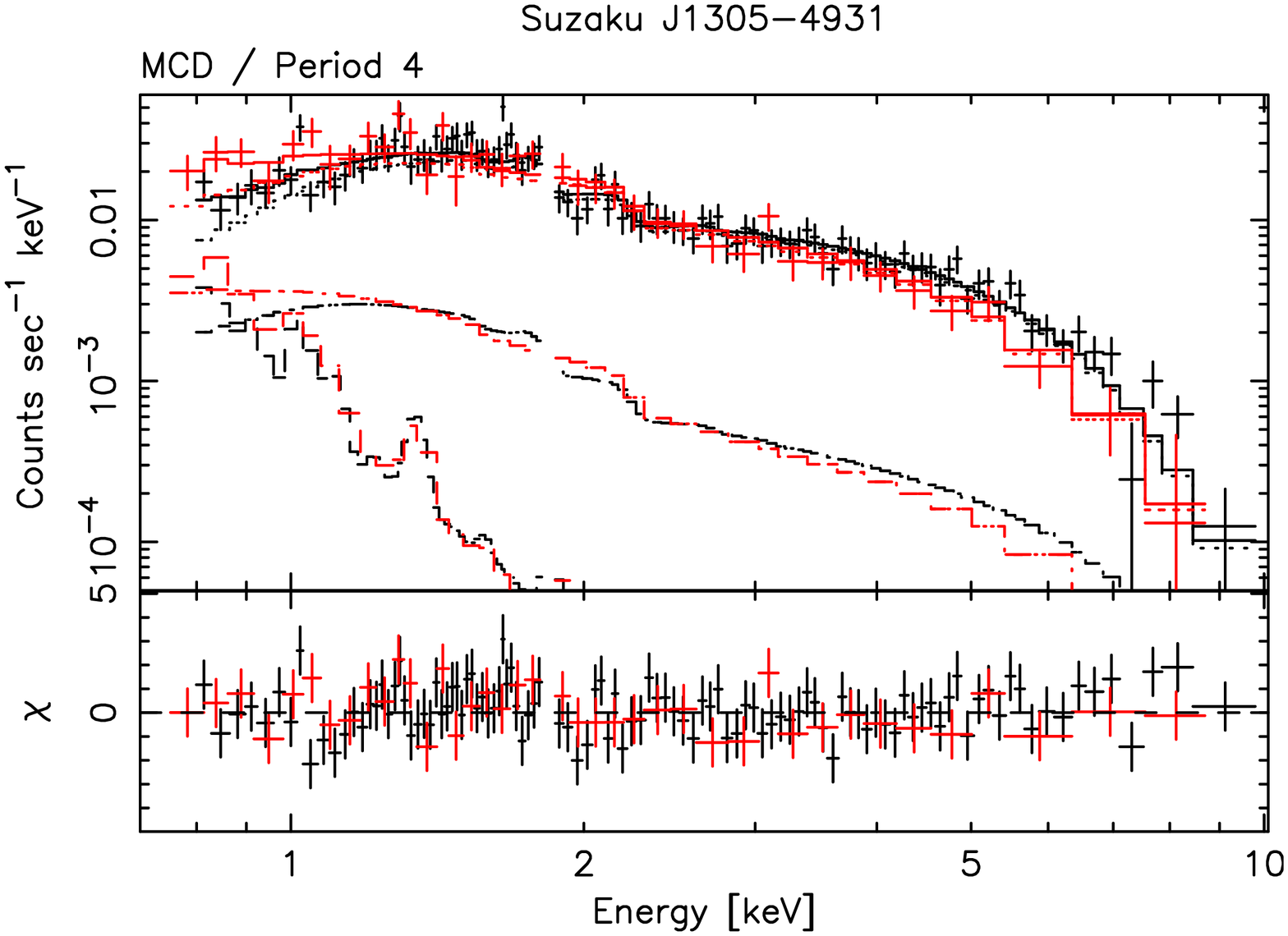}
\end{center}
\caption{XIS spectra of Suzaku J1305-4931, 
obtained in the 4 individual periods 
defined in figure \ref{fig:ulx_lc}.
The top-left, top-right, bottom-left, and bottom-right panel displays
the spectrum of Period 1, 2, 3 and 4, respectively. 
The histograms in each panel show the best-fit MCD model, 
including the emission from the host galaxy. 
}
\label{fig:spec_variability}
\end{figure*}

\begin{figure}[h]
\begin{center}
\FigureFile(80mm,80mm){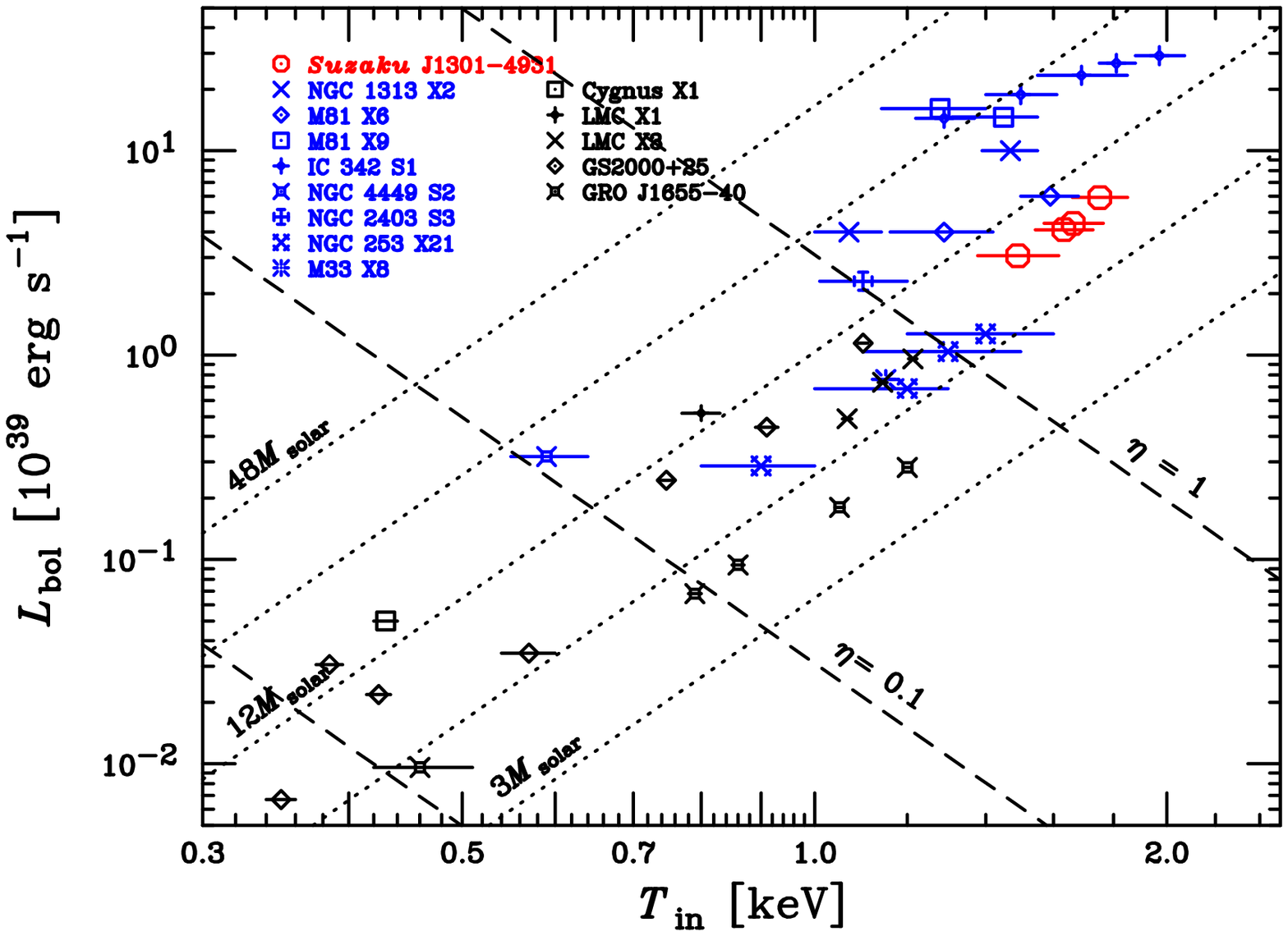}
\end{center}
\caption{
The relation between the bolometric luminosity $L_{\rm bol}$
and the disk temperature $T_{\rm in}$ of Suzaku J1305-4931 (red) 
in comparison with those of well-studied ULXs
or some luminous extragalactic sources 
(blue; \cite{ULX_asca1,NGC2403,ULX_asca2,NGC4449,NGC253X21,M81X9})
and galactic black holes 
(black; \cite{ULX_asca1}, and reference therein). 
The disk inclination of $i = 60^\circ$ is assumed in the plot.
The dotted lines show $L_{\rm bol}$ expected from 
the standard accretion disk, as $L_{\rm bol} \propto T_{\rm in}^4$,
assuming Schwarzschild black holes with several representative masses. 
The dashed lines indicate $\eta = 1$, $0.1$ and $0.01$.
}
\label{fig:Lbol-kT}
\end{figure}

\begin{figure}[h]
\begin{center}
\FigureFile(80mm,80mm){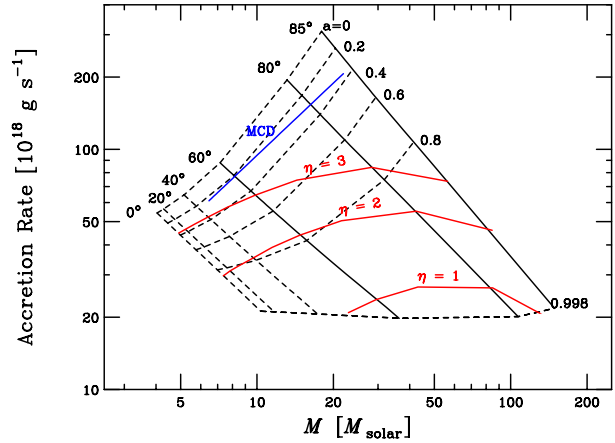}
\end{center}
\caption{The relation between the mass 
$M$ and accretion rate $\dot{M}$,
obtained with the KERRBB model \citep{kerrbb}  
for some representative values of the disk inclination $i$ and
spin parameter $a$.
The solid and dashed lines indicate constant $i$ and $a$, 
respectively. 
The blue line shows the locus of the MCD model. 
The red lines indicate to $\eta = 1$, $2$ and $3$.
}
\label{fig:kerrBB}
\end{figure}

\end{document}